\documentclass[twocolumn,prd,showpacs,nofootinbib,superscriptaddress,preprintnumbers,floatfix]{revtex4-1}

\usepackage{amsmath}
\usepackage{amsfonts}
\usepackage{amssymb}
\usepackage{epsfig}
\usepackage{graphicx}
\usepackage{dsfont}
\usepackage{color}

\usepackage{color}
\usepackage{ulem}

\definecolor{heidelbeer}{rgb}{0,0,0}

\newcommand{\I}{{i}}
\newcommand{\E}{\text{e}}

\newcommand{\fss}[1]{#1\!\!\!/}   
\newcommand{\Eqref}[1]{Eq.~\eqref{#1}}

\newcommand{\Nf}{N_{\mathrm{f}}}
\newcommand{\pat}{\partial_t}
\newcommand{\yb}{\bar{\psi}}

\newcommand{\trho}{\tilde{\rho}}
\newcommand{\hL}{h_\Lambda}
\newcommand{\umf}{U^{\text{MF}}}
\newcommand{\UL}{U_{\Lambda}}
\newcommand{\ML}{M_{\Lambda}}
\newcommand{\mL}{m_\Lambda^2}
\newcommand{\lL}{\lambda_\Lambda}
\newcommand{\mH}{m_{\text{H}}}
\newcommand{\mtop}{m_{\text{top}}}
\newcommand{\Np}{N_{\text{p}}}

\begin{document}

\preprint{}

\title {
Higgs Mass Bounds from Renormalization Flow for a simple Yukawa model} 

\author{Holger Gies}
\email{holger.gies@uni-jena.de}
\affiliation{Theoretisch-Physikalisches Institut, Friedrich-Schiller-Universit\"at Jena, Max-Wien-Platz 1, 
D-07743 Jena, Germany}
\author{Clemens Gneiting}
\email{clemens.gneiting@physik.uni-freiburg.de}
\affiliation{Physikalisches Institut, Albert-Ludwigs-Universit\"at Freiburg, Hermann-Herder-Stra{\ss}e 3,
D-79104 Freiburg, Germany} 
\author{Ren\'{e} Sondenheimer}
\email{rene.sondenheimer@uni-jena.de}
\affiliation{Theoretisch-Physikalisches Institut, Friedrich-Schiller-Universit\"at Jena, Max-Wien-Platz 1, 
D-07743 Jena, Germany}

\begin{abstract}
  We study the functional renormalization group flow of a Higgs-Yukawa
  toy model mimicking the top-Higgs sector of the standard model. This
  approach allows for treating arbitrary bare couplings. For the class
  of standard bare potentials of $\phi^4$-type at a given ultraviolet
  cut-off, we show that a finite infrared Higgs mass range emerges
  naturally from the renormalization group flow itself. Higgs masses
  outside the resulting bounds cannot be connected to any conceivable
  set of bare parameters in this standard-model $\phi^4$ class. By
  contrast, more general bare potentials allow to diminish the lower
  bound considerably. We identify a simple renormalization group
  mechanism for this depletion of the lower bound. If active also in
  the full standard model, Higgs masses smaller than the conventional
  infrared window do not necessarily require new physics at low scales
  or give rise to instability problems.
\end{abstract}


\maketitle

\section{\label{intro} Introduction}

The recent potential discovery of the standard model Higgs boson with
a comparatively low mass of $\mH\simeq 125$GeV \cite{Aad:2012tfa} has
stimulated renewed interest in Higgs mass bounds within the standard
model itself
\cite{Maiani:1977cg,Krasnikov:1978pu,Lindner:1985uk,Altarelli:1994rb,Schrempp:1996fb,Hambye:1996wb}
and beyond
\cite{Cabibbo:1979ay,Espinosa:1991gr,Chen:2012faa,Lebedev:2012zw}. In
particular, arguments based on vacuum stability
\cite{Krive:1976sg,Hung:1979dn,Linde:1979ny,Politzer:1978ic,Sher:1988mj,Lindner:1988ww,Ford:1992mv}
(or sufficient metastability
\cite{Arnold:1989cb,Sher:1993mf,Bergerhoff:1999jj,Isidori:2001bm})
appear to give rise to a lower bound for the Higgs mass
\cite{Ellis:2009tp,EliasMiro:2011aa,Degrassi:2012ry,Alekhin:2012py,Masina:2012tz,Buttazzo:2013uya}. The
measured value for the mass of the discovered scalar boson is either
close to or on top of the bound or might even violate the bound,
depending on various other parameters, most notably the value of the
top mass (in the appropriate scheme) and the strong coupling constant.

The consequences of the true Higgs mass satisfying or violating the
bound can be rather dramatic, ranging from measured constraints on the
underlying UV theory structure, via an upper bound for the scale of new
physics to the prediction of the decay of the universe as we know
it. Therefore, a thorough understanding of Higgs mass bounds within
the standard model is clearly mandatory. 

Even though typical computations of mass bounds are often done with
perturbative (RG-resummed) methods, the problem is generically
nonperturbative. This is obvious for the upper Higgs mass bound -- the
so-called unitarity or triviality bound -- which is, in principle,
related to a strongly coupled Higgs sector in the UV. In perturbation
theory, this becomes manifest from the vicinity to the Landau pole,
indicating the failure of perturbation theory.\footnote{In fact, the
  upper bound is often motivated from the requirement that the
  standard model {\it per definitionem} should be describable within 
  perturbation theory in the UV. Since this is if at all an aesthetic
  but not a physical criterion, we rely on the criterion of triviality
  in the present work.} But also the lower bound involves
nonperturbative information for two reasons: first, the prediction of
infrared (IR) quantities such as Higgs and top masses involve a proper
description of threshold effects. These are nonperturbative, as such
mass scales are related to the couplings. Second, an investigation of
stability issues requires the computation of a full effective potential
for arbitrary field amplitudes.

In a series of works, Higgs mass bounds have therefore recently been studied
within lattice quantum field theory both for a simple $\mathds{Z}_2$
Higgs-Yukawa model \cite{Holland:2003jr,Holland:2004sd,Fodor:2007fn}, as well
as for a Higgs-Yukawa model more similar to and significant for the standard
model \cite{Fodor:2007fn,Gerhold:2007yb}. In particular, the lower Higgs mass
bound arises from the mere criterion of starting from a physically meaningful
bare UV theory on the lattice. No reference to low-energy stability issues had
to be made, and no indications for an instability have been observed. Most
prominently, the simulations of \cite{Gerhold:2010wv,Bulava:2012pb}
essentially rule out or put strong constraints on the existence of a fourth
flavor generation for the measured Higgs boson mass; similar conclusions have
been drawn from analytic considerations \cite{Djouadi:2012ae}.

In the present work, we revisit the Higgs mass bounds by analytic
means using the functional renormalization group (RG). Within a
consistent systematic derivative expansion, the functional RG provides
for a tool to analyze the problem nonperturbatively and allows to
estimate errors of the approximation scheme. In order to concentrate
on the basic mechanisms for the mass bounds, we confine ourselves to
the simple $\mathds{Z}_2$ Higgs-Yukawa model, as it avoids intricate
questions arising from the gauge-Higgs interplay in the full standard
model \cite{Frohlich:1981yi,Maas:2012tj}, while at the same time
maintaining the standard model property that no Goldstone bosons arise
in the broken phase. First functional RG studies of Higgs mass bounds
have already been performed in \cite{Branchina:2005tu,Gneiting:2005}.

In the present work, we particularly concentrate on the influence of
generic UV actions on the Higgs mass bounds. In fact, we find a rather
substantial influence of the precise form of the bare scalar potential
on the lower bound of the Higgs boson. At first sight, this seems to
be at odds with common wisdom of renormalizable field theories that IR
observables should be independent of the details of the microscopic UV
theory. This statement (formulated under suitable mild assumptions)
is, of course, left untouched by our work. However, the main point is
that the notion of a Higgs mass bound is strictly speaking not a pure
IR observable. Higgs mass bounds are typically formulated as a
function of the UV cutoff $\Lambda$, i.e.,
$\mH{}_{,\text{bound}}=\mH{}_{,\text{bound}}(\Lambda)$. Hence, in
order to quantify this dependence, we have to make certain assumptions
about the system at and near the cutoff. This includes the choice of a
regularization scheme, specifying the details of the UV regularization
at the cutoff; in this sense, part of the scheme-dependence of the Higgs
mass bounds is actually physical. And this includes dynamical properties of
the flow near the cutoff which can be rather strongly influenced by
the bare theory. Quantitatively, we find that rather mild
modifications of the bare potential can have a significant impact
on the lower Higgs mass bound. 

This article is organized as follows: in Sect.~\ref{sec:model}, we
briefly introduce our simple toy model. Section \ref{sec:floweq}
summarizes the concepts of the functional RG applied to this model and
presents the resulting flow equations. As a warm-up, a simple
mean-field analysis already illustrating many of the properties of
the Higgs mass bounds is given in
Sect.~\ref{sec:meanfield}. Incidentally, these mean-field properties
do actually not require the functional RG framework, but could equally
well be derived within a large-$N$ type of reasoning. Our main results
based on the nonperturbative RG flow equations are summarized in
Sect.~\ref{sec:flow}.

\section{$\mathds{Z}_2$-symmetric Higgs-Yukawa toy model}
\label{sec:model}

Many of the fluctuation-induced features of Higgs mass bounds in the
standard model can already be studied in a greatly simplified model
involving a Dirac fermion flavor $\psi$ (the top quark) and a real scalar
boson $\phi$. The model is defined by the Euclidean classical action
\begin{equation}
S=\int d^4x \left[ \frac{1}{2}(\partial_\mu \phi)^2 +
  \frac{\bar{m}}{2} \phi^2 + \frac{\bar{\lambda}}{8} \phi^4  
  +  \yb\, \I \fss{\partial} \psi + \I \bar{h}
  \phi\yb\psi \right]. \label{eq:bareaction}
\end{equation}
For later purposes, we allow the top quark to appear in $\Nf$ flavor
copies. We use $\Nf$ merely as an ordering parameter of the
calculation, but not as a physical parameter mimicking the generations
of the standard model. For quantitative statements, we will use
$\Nf=1$. The model is invariant under a
discrete ``chiral'' symmetry,
\begin{equation}
\psi\to \E^{\I \frac{\pi}{2} \gamma_5} \psi, \quad
\yb\to \yb\, \E^{\I \frac{\pi}{2} \gamma_5}, \quad
\phi\to -\phi,\label{eq:discsym}
\end{equation}
which protects the fermions against acquiring a direct mass
term. Since the symmetry is only discrete, its spontaneous breaking
owing to a nonzero expectation value for the scalar field
$v=\langle\phi\rangle$ does not give rise to massless Goldstone
bosons. This feature mimics the property of the standard model that
the Goldstone modes are eaten by the massive electroweak gauge bosons.

The quantum theory corresponding to \Eqref{eq:bareaction} has to be
defined with a finite ultraviolet (UV) cutoff $\Lambda$ which, together
with a specified regularization prescription, remains an implicit
physical parameter of the theory. This is because triviality inhibits
an ultraviolet extension to arbitrarily high scales while keeping the
physical low-energy parameters fixed \cite{Wilson:1973jj}; in
perturbation theory, this feature is reflected by the existence of a
Landau pole in the running coupling.

Solving the so defined quantum theory provides for a mapping from the
microscopic bare parameters $\bar{m}^2,\bar{\lambda}, \bar{h},
\Lambda$ and possibly further RG irrelevant bare couplings to the
set of physical parameters which are given by the top mass $\mtop$,
the Higgs mass $\mH$, the vacuum expectation value $v$ and still the
cutoff $\Lambda$. These physical parameters are directly  related to
renormalized couplings in the quantum effective action, such as the
renormalized Yukawa coupling $h$, see below, and the effective potential
$U(\rho)$, where $\rho=\frac{\phi^2}{2}$. Denoting the minimum of the
effective potential by $\rho_0$, we identify
\begin{equation}
v={Z_\phi^{1/2}}\langle\phi\rangle=\sqrt{2{Z_\phi}\rho_0}, 
\,\, \mtop^2={v^2} h^2, \,\,
\mH^2={v^2 \frac{U''(\rho_0)}{Z_\phi^2}} , \label{eq:physpar} 
\end{equation}
where the wave function renormalization $Z_\phi$ is introduced below.
In this work, we consider the vacuum expectation value and the top
mass as given, $v\simeq246$GeV and $\mtop\simeq 173$GeV.\footnote{We
  use here the value for the top mass measured by kinematically
  reconstructing its decay products and comparing these to Monte Carlo
  simulations. For Higgs mass bounds, actually the pole mass is
  considered to be the appropriate quantity, which could significantly
  differ from the experimentally quoted value
  \cite{Alekhin:2012py}. As a rule of thumb, an uncertainty of $\sim
  1$GeV in the top mass leads to a $\pm2$GeV variation of the lower
  Higgs mass bound for large cutoffs $\Lambda$. In any case,
  quantitative results of the present toy model should anyway only be
  considered as an illustrative example.} Furthermore, choosing a
fixed cutoff $\Lambda$ leaves only $\mH$ as a free parameter which
becomes a function of the whole set of microscopic bare parameters.

Constraints on the Higgs mass are now obtained if the region of
attainable Higgs masses is bounded for any given combination of bare
parameters. These bare parameters are essentially unconstrained, as
they are provided by a yet unknown underlying microscopic theory (UV
completion). Only a stable bare scalar potential bounded from below is
required in order to facilitate a meaningful definition of the quantum
theory. In the present work, we start with the standard class of
initial bare $\bar\lambda \phi^4$ potentials. UV stability then
implies that $\bar\lambda\geq 0$ for this class of potentials. We then
extend our considerations to more general potentials. For
instance, also a negative $\bar\lambda$ is permitted if the potential
is stabilized for large $\phi$, e.g., by positive $\phi^6,\phi^8,
\dots$ terms in the bare potential. We emphasize that these
higher-order terms cannot be excluded by referring to
renormalizability criteria. This is because we consider them to be
present in the microscopic UV potential at a fixed (possibly physical)
UV cutoff $\Lambda$. Presently no experiment can impose relevant
constraints on such terms which could arise from an underlying UV
completion of the standard model. Renormalizability rather tells us
that the IR is dominated by the power-counting ``renormalizable''
operators in the standard model, provided that the UV theory starts
near the perturbative Gau\ss ian fixed point.

\section{Renormalization Flow}
\label{sec:floweq}

As an alternative to the functional-integral definition of continuum
quantum field theory, we use a differential formulation provided by
the functional RG. A convenient version is given by the flow equation
for the effective average action $\Gamma_k$, which interpolates
between the bare action $\Gamma_{k=\Lambda}= S$ at the UV cutoff
$\Lambda$ and the full quantum effective action $\Gamma=\Gamma_{k=0}$
\cite{Wetterich:1992yh}. The latter corresponds to the generator of 
fully-dressed proper vertices. The variation of the effective action
with respect to the scale $k$ is given by the Wetterich equation
\begin{equation}
k\,\partial_{k}\Gamma_{k}
\equiv\partial_{t}\Gamma_{k}
=\frac{1}{2}\mathrm{STr} \left[(\partial_{t}R_{k})
  (\Gamma_{k}^{(2)} +R_{k})^{-1} \right]
, \,\, t=\ln\frac{ k}{\Lambda}.
\label{eq:flow_eq1}
\end{equation}  
Here, $\Gamma_{k}^{(2)}$ denotes the second functional derivative with respect
to the fluctuating fields $\Phi=(\phi,\psi,\yb)$, and the super-trace also
includes a minus sign for the fermions.  The regulator $R_{k}$ in the
denominator is chosen such that it suppresses IR modes below the scale $k$, and its
derivative $k\partial_{k}R_{k}$ establishes ultraviolet (UV) finiteness; as a
consequence, the flow of $\Gamma_k$ is dominated by fluctuations with momenta
$p^2\simeq k^2$, implementing the concept of smooth momentum-shell
integrations, for reviews see~\cite{Berges:2000ew,Aoki:2000wm,Pawlowski:2005xe,Gies:2006wv,
  Delamotte:2007pf,Kopietz:2010zz,Braun:2011pp}.

As we are working with an explicit finite cutoff $\Lambda$, also the
choice of the regularization scheme strictly speaking belongs to the
definition of the model. This scheme is here specified in terms of the
regulator function $R_{k}$, more precisely in terms of the regulator
shape functions $r(p^2/k^2)$, $r_{\text{F}}(p^2/k^2)$ introduced in
the appendix. From the viewpoint of the model definition, these shape
functions determine how the modes are physically cut off in the
UV. Since a change of the regularization scheme such as a change of
the shape functions can be mapped onto a change of the initial
conditions for the bare couplings, we keep the regulator fixed in the
present work and vary the bare couplings.

In addition to perturbative expansions, nonperturbative approximation schemes
can be devised for the flow equation. Systematic and consistent
expansion schemes which do not rely on a perturbative coupling ordering are,
for instance, the vertex expansion or the derivative expansion. 

In this work, we study the renormalization flow of the Yukawa system
nonperturbatively within the following truncation based on the derivative
expansion:
\begin{equation}
\Gamma_{k}  =  \int_x \left[
 \frac{Z_{\phi,k}}{2} \, ( \partial_{\mu} \phi )^2 +  U_{k}(\rho)
 + Z_{\psi,k} \, \yb\, \I \fss{\partial}
  \psi + i \, \bar{h}_{k} \, \phi  \yb \psi \right],  \label{eq:trunc}
\end{equation}
where $\rho=\frac{1}{2} \phi^2$, and the potential $U_k$ generally
includes arbitrary powers of the field. 
%
%
In fact, the accuracy of the derivative expansion for scalar theories
has been verified quantitatively in many contexts. Here, we actively
study its convergence by comparing leading-order (LO) results (obtained for
$Z_{\phi,k}=1, Z_{\psi,k}=1$) to next-to-leading order (NLO) results. We find no
signatures of a failure of this expansion even at comparatively strong
coupling, see below.

Inserting this {\it ansatz} \eqref{eq:trunc} into the flow equation
\eqref{eq:flow_eq1} provides us with the RG flows of $\bar{h}_{k},U_k$ and the
wave function renormalizations $Z_{\phi,k}$ and $Z_{\psi,k}$; the latter flows
will be followed in terms of the anomalous dimensions
\begin{equation}
\eta_\phi=-\pat \ln Z_{\phi,k}, \quad \eta_\psi=-\pat \ln
Z_{\psi,k}. \label{eq:etadef} 
\end{equation}
The flow equation for the effective potential reads
\begin{eqnarray}
\partial_{t} U_{k} & = & 2 \, v_{d} \, k^{d} \, \Big[ l_{0}^{d} \left(
                   k^{-2} \, Z_{\phi,k}^{-1} \, \left[ 2 \, \rho \,
                   U_{k}'' + U_{k}' \right];\eta_{\phi} \right)
                   \nonumber \\ 
                   &   & \qquad - \Nf d_{\gamma} \, l_{0}^{(F)\,d} \left(
                   2 \, k^{-2} \, Z_{\psi,k}^{-2} \,
                   \bar{h}_{k}^{2} \, \rho;\eta_{\psi} \right)
                   \Big],  \label{eq:potflow}
\end{eqnarray}
where the primes denote derivatives with respect to $\rho$, and
$v_{d}^{-1} = 2^{d+1} \, \pi^{d/2} \, \Gamma(d/2)$. For generality, we
work in $d$ dimensions and with a $d_\gamma$ dimensional
representation of the Dirac algebra. We will later specialize to $d=4$
and $d_\gamma=4$. The threshold functions $l_{0}^{d}$ and
$l_{0}^{(F)\,d}$ {arise from the integration over the
  loop momentum and carry the non-universal regulator dependence. For
  any physically admissible regulator, they} approach finite constants
for vanishing argument and decrease to zero for large first argument,
describing the decoupling of massive modes; details can be found in
Appendix \ref{appA}.
 
It is useful to introduce renormalized fields
\begin{equation}
\tilde{\phi} =Z_{\phi,k}^{1/2} \phi, \quad
\tilde{\psi} = Z_{\psi,k}^{1/2} \psi, \label{eq:renfields}
\end{equation}
as well as the dimensionless renormalized Z${}_2$ invariant quantity
\begin{equation}
\trho= Z_{\phi,k} \, k^{\, 2-d} \rho. \label{eq:trhodef}
\end{equation}
The dimensionless renormalized Yukawa coupling is defined by
\begin{equation}
h_{k}^{2} = Z_{\phi,k}^{-1} \, Z_{\psi,k}^{-2} \, k^{\, d-4} \,
\bar{h}_{k}^{2}, \label{eq:hqdef}
\end{equation}
and the dimensionless potential simply is:
\begin{equation}
u_{k} = U_{k} \, k^{\, -d}. \label{eq:potdef}
\end{equation}
The flow of $u_k$ for fixed $\trho$ is given by

\begin{widetext}

\begin{eqnarray} 
  \partial_{t} \, u_{k} 
  & = & -d \, u_{k} + (d-2+\eta_{\phi}) \,
  \trho \, u'_{k} + 2 \, v_{d} \, \Big[ l_{0}^{d}\left(u_{k}' + 2 \,
    \trho  \, u_{k}''; \eta_{\phi} \right)
    - \Nf d_{\gamma} \, l_{0}^{(F)\,d}\left(2 \, \trho \,
    h_{k}^{2} ; \eta_{\psi} \right) \Big],
  \label{PotentialFlowEquation} 
\end{eqnarray}
where primes now denote derivatives with respect to $\trho$. The flow
of the Yukawa coupling is of the form
\begin{eqnarray}
\partial_{t} \, h_{k}^{2} 
& = & \left[ \eta_{\phi} + 2 \, \eta_{\psi} + d - 4 \right] \,
h_{k}^{2}
 + 8 \, h_{k}^{4} \,v_{d} \,
l_{1,1}^{(FB)\,d}\left(\omega_{1},\omega_{2};
     \eta_{\psi},\eta_{\phi}\right)  \label{eq:flowhq} \\ 
&& - \left[ 48 \, \kappa_{k} \, u_{k}''(\kappa_{k}) 
  + 32 \, \kappa_{k}^{2} \, u_{k}'''(\kappa_{k}) \right] 
\, h_{k}^{4} \, v_{d} \, l_{1,2}^{(FB)\,d}\left(\omega_{1},\omega_{2};
                      \eta_{\psi},\eta_{\phi} \right) 
 - 32 \, h_{k}^{6} \, \kappa_{k} \, v_{d} \, 
l_{2,1}^{(FB)\,d}\left(\omega_{1},\omega_{2};
   \eta_{\psi},\eta_{\phi}\right), \nonumber 
\end{eqnarray}
with
\begin{eqnarray}
\omega_{1} & = & 2 \, \kappa_{k} \, h_{k}^{2}, \quad
\omega_{2}  =  u_{k}'(\kappa_{k}) + 2 \kappa_{k} u_{k}''(\kappa_{k})
\, , \label{eq:omegadef}  
\end{eqnarray}
and $\kappa_k=\trho_{\text{min}}$ denotes the minimum of the
potential; i.e., if $\kappa_k\neq 0$ then $u_k'(\kappa_k)=0$. Finally,
the anomalous dimensions are determined by
\begin{eqnarray}
\eta_{\phi} & = & 8 \, \frac{v_{d}}{d} 
\bigg[ \kappa_{k} \left[3 \, u_{k}''(\kappa_{k})+2 \, \kappa_{k} \,
    u_{k}'''(\kappa_{k}) \right]^{2} m_{4,0}^{d}\left(2 \, \kappa_{k}
  \, u_{k}''(\kappa_{k})+u_{k}'(\kappa_{k}),0;\eta_{\phi}\right)
  \nonumber \\ 
&   & \qquad\;\;\; + \Nf d_{\gamma} \, h_{k}^{2} \Big[
    m_{4}^{(F)\,d}\left(2 \, \kappa_{k} \,
    h_{k}^{2};\eta_{\psi}\right)  
-2 \, \kappa_{k} \, h_{k}^{2} \, m_{2}^{(F)\,d}\left(2 \, \kappa_{k}
\, h_{k}^{2};\eta_{\psi}\right) \Big] \bigg], \label{eq:etaphi}\\
\eta_{\psi} &=& 8 \, h_{k}^{2} \, \frac{v_{d}}{d} \,
m_{1,2}^{(FB)\,d}\left(2 \, \kappa_{k} \, h_{k}^{2},2 \, \kappa_{k} \,
u_{k}''(\kappa_{k}) +
u_{k}'(\kappa_{k});\eta_{\psi},\eta_{\phi}\right), \label{eq:etapsi}
\end{eqnarray}
where the threshold functions are again discussed in Appendix
\ref{appA}. These flow equations can be compared to those of similar
investigations in the literature
\cite{Hofling:2002hj,Gies:2009hq,Braun:2010tt} within different
physical contexts. Once the flow equations have been solved for
suitable initial conditions, we can read off the fully renormalized
long-range quantities in the limit $k\to0$. For instance, the physical
quantities defined in \Eqref{eq:physpar} require the renormalized
Yukawa coupling $h=h_{k\to0}$ and the wave function renormalization
$Z_\phi=Z_{\phi,k\to0}$. The renormalized vacuum expectation value is
obtained from $v^2=\lim_{k\to0} 2k^2 \kappa_k$.

\end{widetext}

\section{Mean-field analysis}
\label{sec:meanfield}

Let us first perform a mean-field analysis, corresponding to a
one-loop approximation of the effective potential including fermion as
well as boson fluctuations, while keeping the wave function renormalizations
and the Yukawa coupling fixed,
\begin{equation}
Z_{\phi,k},Z_{\psi,k}\to 1, \quad h_k \to \hL.
\end{equation}
The mean-field effective potential $\umf$ could, of course, be
calculated directly from a Gau\ss ian approximation of the generating
functional, yielding the standard $\log\det$-formula. Nevertheless, we
derive it from the flow equation, since it provides direct access to
the use of an arbitrarily shaped regulator function, which can be used
to model the physical UV cutoff mechanism.

The standard mean-field (MF) approximation is equivalent to the large-$\Nf$
approximation, taking only fermionic fluctuations into account. The
corresponding mean-field effective potential is obtained from the flow
equation \eqref{eq:potflow} by integrating the fermion contributions $\sim\Nf$
from $k=\Lambda$ to $0$, while keeping the potential on the right-hand side
fixed at $U_k \to \UL$.  We obtain for the mean-field effective potential
  \begin{eqnarray}
    \umf_k(\rho) &=& \UL(\rho)    \label{eq:MFU}\\
&& +\frac{\Nf d_\gamma}{2} \int_p \ln
    \left(
      \frac{p^2(1+r_{\text{F}}(p^2/\Lambda^2))^2 +2\bar{h}_{\Lambda}^2\rho}
           {p^2(1+r_{\text{F}}(p^2/k^2))^2 +2\bar{h}_{\Lambda}^2\rho}
    \right),
    \nonumber
  \end{eqnarray}
where $\int_p= \int \frac{d^d p}{(2\pi)^d}$.  The extended mean-field
(EMF) approximation is obtained by including also the scalar fluctuations on
the same Gau\ss ian level. Introducing the abbreviation 
\begin{equation}
\ML^2(\rho)=\UL'(\rho)+2\rho\UL''(\rho),
\end{equation}
we find,
  \begin{eqnarray}
    U^{\text{EMF}}_k(\rho) &=& \umf_k(\rho)    \label{eq:EMFU}\\
   && -\frac{1}{2} \int_p \ln
    \left[
      \frac{p^2(1+r(p^2/\Lambda^2)) + \ML^2(\rho)}
           {p^2(1+r(p^2/k^2)) + \ML^2(\rho)}
    \right].
 \nonumber
  \end{eqnarray}
Whereas the mean-field approximation becomes exact in the strict
large-$\Nf$ limit, no such anchoring to an exact limit is known for
the extended-mean-field approximation. Moreover, further subtleties
arise in the extended-mean-field case from convexity violations and
complex solutions for the potential as discussed in
\cite{Weinberg:1987vp}. These subtleties of the extended mean-field
approximation are however irrelevant for the nonperturbative functional
RG solution discussed below. Hence, we will mainly stay within the
standard mean-field approximation in the following for the purpose of
illustration. 

For both approximations, the momentum integration can be done
analytically for a suitable choice of the regulator shape functions
$r(x),r_{\text{F}}(x)$. For instance, for the linear regulator
(cf. App.~\ref{appA}) we obtain in the limit $k\to0$
and in $d=4$ dimensions (where $\bar{h}_{\Lambda}=h_{\Lambda}$)
\begin{widetext}
\begin{eqnarray}
U^{\text{EMF}}(\rho)&=&\UL(\rho) + \frac{1}{64 \pi^2} \Bigg\{ [\ML^2(\rho) -\ML^2(0)
  -2\Nf d_\gamma \hL^2\rho] \Lambda^2  +4\Nf d_\gamma \hL^4 \rho^2 \ln
\frac{\Lambda^2 +2\hL^2\rho}{2\hL^2\rho} \nonumber\\
&&\qquad\qquad \qquad\qquad 
-\ML^4(\rho) \ln \frac{\Lambda^2+\ML^2(\rho)}{\ML^2(\rho)} 
+\ML^4(0) \ln \frac{\Lambda^2+\ML^2(0)}{\ML^2(0)} \Bigg\},
\label{eq:umf}
\end{eqnarray}
\end{widetext}
where we have normalized $U^{\text{EMF}}(\rho)$ such that
$U^{\text{EMF}}(0)=0$.  In the following we will show that \Eqref{eq:umf} can
be used to illustrate the appearance of a lower bound for the Higgs mass.

\subsection{Bare potentials of $\phi^4$-type}

Let us confine ourselves to bare potentials of
$\phi^4$-type,
\begin{equation}
\UL(\rho)=\mL \rho + \frac{\lL}{2} \rho^2. \label{eq:conf}
\end{equation}
For a given UV cutoff $\Lambda$, two out of the three bare parameters
$\mL,\lL,\hL$ can be fixed by fixing the top mass and the vacuum
expectation value; more precisely, fixing $\hL=\mtop/v$ and
determining $\mL$ from the transcendental equation
\begin{equation}
  U^{\text{EMF}}{}'(\rho_0=v^2/2)=0,\label{eq:umft}
\end{equation}
leaves us with the Higgs mass as a function of the bare scalar
coupling, $\mH=\mH(\lL)$. In the standard mean-field approximation, it
is easy to see that $\mH=\mH(\lL)$ increases monotonically with $\lL$,
therefore a lower bound on the Higgs mass is obtained from the lowest
possible value of $\lL$, which is $\lambda_{\Lambda,\text{min}}=0$ for
potentials of the form of \Eqref{eq:conf}. (In the extended mean-field
approximation, the same conclusion holds unless $\lL$ approaches the 
strong-coupling value $\lL\to \frac{8}{3}h_{\Lambda}^2$ where an
EMF-artifact induces singular behavior).

Equation~\eqref{eq:umft} can easily be solved numerically. For an
analytical estimate, let us stay within the mean-field approximation and keep
only the terms $\sim\Nf$. Determining $\mL$ from the condition
$U^{\text{MF}}{}'(\rho_0=v^2/2)=0$ for fixed values of $\mtop$ and
$v$, we find (setting $\Nf=1$, $d_\gamma=4$)
\begin{eqnarray}
\mL(\Lambda,\lL)&=&- \frac{\lL}{2} v^2+\frac{\hL^2}{8\pi^2} \Lambda^2
\label{eq:MFmL}\\
&&- \frac{\hL^4
  v^2}{8\pi^2} \left[2\ln\left( 1+ \frac{\Lambda^2}{\mtop^2} \right) -
  \frac{\Lambda^2}{\Lambda^2 + \mtop^2} \right] . 
\nonumber
\end{eqnarray}
This fixes the effective mean-field potential as a function of $\lL$ and
$\Lambda$, yielding the Higgs mass
\begin{eqnarray}
\mH^2(\Lambda,\lL)&=&  v^2 U^{\text{MF}}{}''(v^2/2) \nonumber\\
&=&\frac{\mtop^4}{4\pi^2 v^2}\! \left[2 \ln \left(\! 1+
    \frac{\Lambda^2}{\mtop^2} \right)\! - \frac{
      3\Lambda^4+2\mtop^2\Lambda^2}{(\Lambda^2+\mtop^2)^2}
    \!\right]\nonumber\\
&& + v^2 \lL. \label{eq:MFmH}
\end{eqnarray}
This renders explicit that the lower bound for UV
potentials of the form of \Eqref{eq:conf} is given by
$\mH(\Lambda,\lL=0)$.  

\begin{figure}
{\centering
\includegraphics[width=8.5cm]{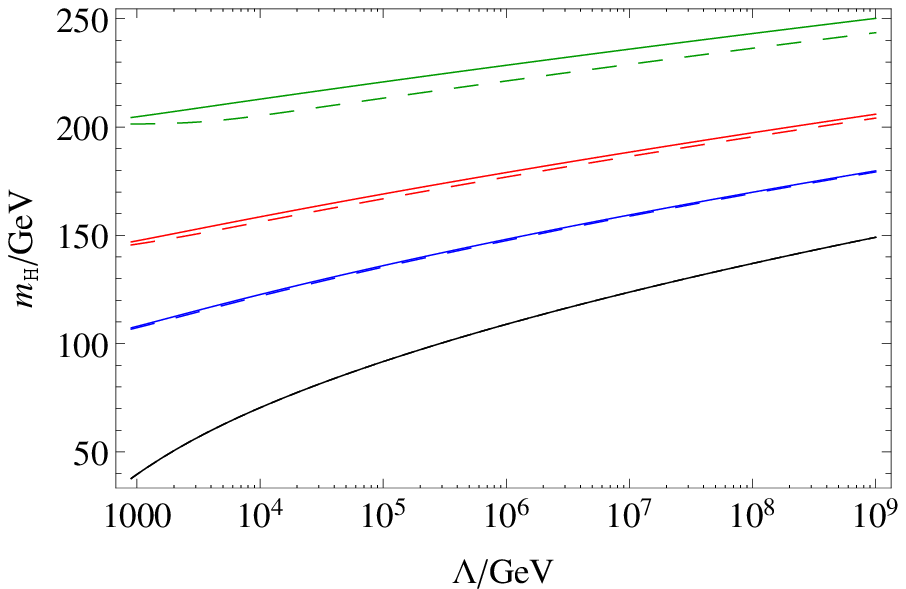}
} 
\caption{Extended mean-field analysis of the lower bound for the Higgs
  mass $\mH$ versus the UV cutoff $\Lambda$, based on a bare potential
  $\UL$ of $\phi^4$-type for $\Nf=1$. For an initial potential which
  is flat apart from a mass term $\UL=\frac{1}{2} \mL \phi^2$, the
  fermionic fluctuations drive the Higgs mass to a finite minimal
  value. The solid lines correspond to standard mean-field theory
  accounting only for top fluctuations, cf. \Eqref{eq:MFmH}, whereas
  the dashed lines also include scalar fluctuations on the Gau\ss ian
  level (extended mean-field). The four different line sets correspond
  to increasing values of the initial $\phi^4$ coupling of
  $\lL=0,\frac{1}{6},\frac{1}{3},\frac{2}{3}$ from bottom to top.}
\label{fig:mHiggsMF}
\end{figure}

The mean-field analysis performed here gives a first insight into how
lower bounds for the Higgs mass follow from the mapping from bare to
renormalized quantities. It also exemplifies that the mere existence
of a lower bound on the Higgs mass for bare potentials of
$\phi^4$-type is essentially a consequence of top fluctuations that
drive the curvature of the effective potential at its nontrivial
minimum to finite values. This statement will also hold on the
nonperturbative level. We plot the mean-field results for the Higgs
mass as a function of $\Lambda$ for various values of $\lL$ in
Fig.~\ref{fig:mHiggsMF} as solid lines. 

The plot also shows corrections from bosonic fluctuations as described by
extended mean-field theory $U^{\text{EMF}}(\rho)$ as dashed lines for the same
values of $\lL$. We observe that scalar fluctuations tend to decrease the
Higgs mass values. This agrees with the fact that scalar fluctuations drive
the effective potential towards the symmetric regime, thus depleting also the
curvature near the minimum. However, the lower bound of the Higgs mass remains
unaffected by the scalar fluctuations, because the scalar field is
non-interacting for $\lL=0$ in the EMF approximation.

\subsection{Generalized bare potentials}

The lower Higgs mass bound determined above arises from the fact that
the values for the bare quartic coupling $\lambda_{\Lambda}$ are
bounded from below. This is necessary in order to start with a
well-defined theory in the UV for our confined bare potentials
\eqref{eq:conf} of $\phi^4$-type.  Such a restriction on the bare
potential is typically also required in perturbation theory because
higher-order operators are perturbatively non-renormalizable. By
contrast, the Wetterich equation provides us with a nonperturbative
tool, so we can study also the influence of RG irrelevant higher-order
operators on the flow of the effective average
action. {Alternatively, this could also be studied
  with perturbative methods in an effective-theory approach.}

In the following we address the question how modifications $\Delta
U_{\Lambda}(\rho)$ of the quartic bare potential can exert an influence on the
lower Higgs mass bound. The bare potential can in principle be an arbitrary
function of the scalar field. The only constraint which we impose is that the
potential is bounded from below in order to start from a well-defined quantum
field theory at the cutoff. We emphasize that no further experimental
constraints exist. The simplest extention of the standard potential has an
additional operator of the form $\phi^6$.
\begin{align}
U_{\Lambda}(\rho) &= m_{\Lambda}^2\rho+\frac{\lambda_{\Lambda}}{2}\rho^2 + \Delta U_{\Lambda}(\rho)\notag\\
&= m_{\Lambda}^2\rho + \frac{\lambda_{\Lambda}}{2}\rho^2 + \frac{\lambda_{3,\Lambda}}{6\Lambda^2}\rho^3.
\end{align}
Again, in the mean-field case \Eqref{eq:umft} can be solved explicitly for $m_{\Lambda}^2$, 
yielding the Higgs mass as a function of $\lambda_{\Lambda}$ and $\lambda_{3,\Lambda}$ for a given 
cutoff, $\mH=\mH(\lambda_{\Lambda},\lambda_{3,\Lambda})$. With $\lambda_{3,\Lambda}$ positive 
we can study a wider range of values for the bare quartic coupling. The Higgs mass reads
\begin{widetext}
\begin{align}
\mH^2(\Lambda,\lambda_{\Lambda},\lambda_{3,\Lambda}) &= \frac{m_{\mathrm{top}}^4}{4\pi^2v^2} \left[ 2\ln{\left(1+\frac{\Lambda^2}{m_{\mathrm{top}}^2}\right)} - \frac{3\Lambda^4+2m_{\mathrm{top}}^2\Lambda^2}{(\Lambda^2+m_{\mathrm{top}}^2)^2} \right] + v^2\lambda_{\Lambda} + \frac{v^4}{2\Lambda^2}\lambda_{3,\Lambda}.
\label{eq:mHmfep}
\end{align}
\end{widetext}
Obviously, we are able to construct a theory with a Higgs mass below the
previous lower bound if the contribution of the term $\sim \lL$ for $\lL<0$
exceeds that of the positive term $\sim \lambda_{3,\Lambda}$.

\begin{figure}
{\centering
\includegraphics[width=8.5cm]{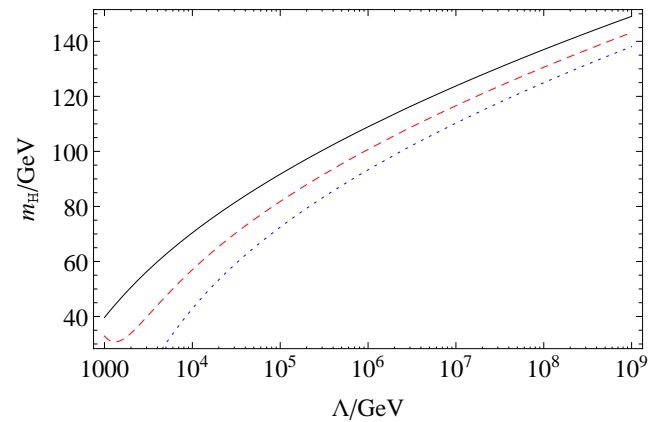}
} 
\caption{Extended mean-field analysis of the lower bound for the Higgs mass
  $\mH$ versus the UV cutoff $\Lambda$, based on a bare potential $\UL$ of
  $\phi^6$-type for $\Nf=1$. We have plotted the lower bound in the $\phi^4$
  theory ($\lL=\lambda_{3,\Lambda}=0$) as solid black line. Theories with bare
  couplings $\lL=-\frac{1}{30}$ and $\lambda_{3,\Lambda}=\frac{2}{3}$ are
  depicted as red dashed line, and $\lL=-\frac{1}{15}$ and
  $\lambda_{3,\Lambda}=2$ as blue dotted line.}
\label{fig:mHiggsMF2}
\end{figure}

The same mechanism works in the extended mean-field analysis but
there it requires a solution to the transcendental \Eqref{eq:umft} in order to
determine $m_{\Lambda}^2$.  A numerical solution is plotted in
Fig.~\ref{fig:mHiggsMF2} for different values of $\lambda_{\Lambda}$
and $\lambda_{3,\Lambda}$. Furthermore, we have checked that for the
given masses no additional minimum appears in the effective potential
besides the one at $v=246\,$GeV.

Let us finally remark that upper bounds cannot meaningfully be studied
in the mean-field approximation; this is because ``RG improvement'' is
necessary to observe the nonperturbative approach to triviality (reflected by
the Landau-pole behavior within RG-improved perturbation theory).

\section{Nonperturbative Higgs mass bounds}
\label{sec:flow}

The mean-field approximation has turned out to be remarkably accurate by
direct comparison with nonperturbative lattice simulations for the present
model \cite{Holland:2003jr,Holland:2004sd}. As lattice simulations are
typically limited as far as the separation of the UV scale from the physical
scales is concerned, a nonperturbative continuum analysis of beyond mean-field
theory seems indispensable in order to appropriately account for scalar
fluctuations and the mutual back-reactions between fermionic and scalar
fluctuations on a wide range of scales.

For the solution of the flow equations, we use the formulation in terms
of dimensionless renormalized quantities as introduced in
Sect.~\ref{sec:floweq}. To leading-order in the derivative expansion,
we solve the flow equations for the effective potential
$u_k$ and for the Yukawa coupling $h_k$. At next-to-leading order, we
include the wave-function renormalizations $\eta_\phi$ and
$\eta_\psi$. 

Since we are mainly interested in the properties of the effective
potential near its minimum, we use a polynomial expansion of the
potential. The stability and convergence of this expansion will be
checked explicitly. In the symmetric regime (SYM) where the minimum of
the potential occurs at $\kappa_k=0$,
we use the truncated expansion
\begin{equation}
u_k=\sum_{n=1}^{\Np} \frac{\lambda_n}{n!} \, \trho^n, \label{eq:usym}
\end{equation}
such that the potential is parameterized by $\Np$ couplings
$\lambda_n$ (the mass term is related to $\lambda_1$ and we identify
the $\phi^4$ interaction as $\lambda\equiv\lambda_2$). In
the symmetry-broken regime (SSB), we instead use
\begin{equation}
u_k=\sum_{n=2}^{\Np} \frac{\lambda_n}{n!} \, (\trho-\kappa_k)^n. \label{eq:ussb}
\end{equation}
The flows of $\lambda_1, \dots, \lambda_{\Np}$ (SYM), or $\kappa_k,\lambda_2, \dots
\lambda_{\Np}$ (SSB), can directly be derived from
\Eqref{PotentialFlowEquation}. 

For small bare scalar coupling $\lL\equiv \lambda_{2,\Lambda}$, a physical
flow typically starts in the SYM regime. Near the electroweak scale, fermionic
fluctuations drive the system into the SSB regime at a scale $k_{\text{SSB}}$,
where we have to switch from the SYM flow to the SSB flow. Here, a nonzero
minimum builds up, inducing masses for the fermions and the Higgs scalar. This
leads to a decoupling of the modes, and the flow freezes out completely; i.e.,
all right-hand sides of the flow equations go to zero for $k\to 0$. For large
bare scalar coupling $\lL$, the physical flow starts already in the SSB regime
with a small value for $\kappa_k$. The flow can still run over many scales
until $\kappa_k$ grows large near the electroweak scale, implying again the
decoupling of all modes.

\subsection{$\phi^4$-type bare potentials}
Let us again start with the restricted class of bare potentials of
$\phi^4$-type, 
\begin{equation}
u_\Lambda=\lambda_{1,\Lambda} \trho +
\frac{\lambda_{\Lambda}}{2} \trho^2, \label{eq:conf2}
\end{equation}
where $\lambda_{1,\Lambda}\equiv\mL/\Lambda^2$ for a wave function
renormalization $Z_{\phi,\Lambda}=1$. For a given cutoff $\Lambda$,
the flow equations map the bare parameters $\mL$, $\lL$, $\hL$ onto
the physical parameters $v$, $\mtop$, $\mH$. In practice, we tune
$\mL$ to establish the correct vacuum expectation value
$v\simeq246$GeV for a given cutoff $\Lambda$. This is, in fact, a
fine-tuning problem, corresponding to the problem of separating the
scale hierarchies in the standard model.  At the same time, $\hL$ is
varied until the flow ends at the right value of $\mtop$. This leaves
us with the Higgs mass as a function of $\lL$ for a given cutoff
$\Lambda$, $\mH=\mH(\Lambda,\lL)$, where $\lL$ is allowed to be an a
priori arbitrary non-negative real number for the class of bare
potentials \eqref{eq:conf2}.

\begin{figure}
{
\centering\includegraphics[width=8.5cm]{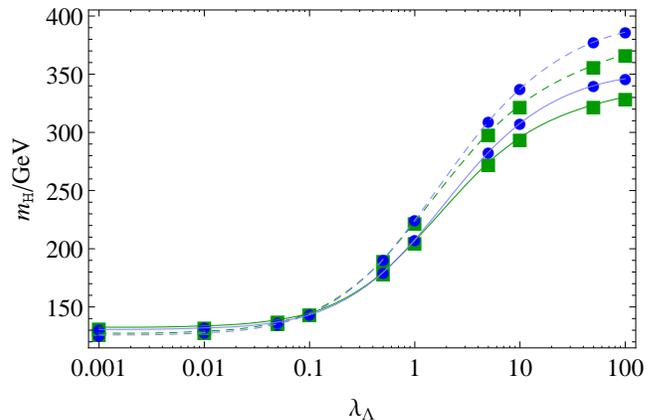}
} 
\caption{Higgs mass values versus the bare scalar coupling $\lL$ for a
  cutoff $\Lambda=10^7$GeV. The dashed lines denote the results within
  LO derivative expansion; the NLO deviates from the LO result by at
  most 10 \% for large coupling, demonstrating the satisfactory
  convergence of the derivative expansion. Also the convergence of the
  polynomial expansion is shown: green lines with squares are obtained
  within the lowest nontrivial order with $\Np=2$, blue lines with
  circles denote the $\Np=4$ result; even higher orders $N_P=6,8$ show no
  further deviation from the $\Np=4$ curves. }
\label{fig:lambdamap}
\end{figure}

In Fig.~\ref{fig:lambdamap}, we depict this function $\mH(\lL)$ for a
cutoff $\Lambda=10^7$GeV for various approximations. For $\lL\lesssim
0.01$, the Higgs mass becomes rather independent of $\lL$ approaching
its lower bound. {This observation is in perfect agreement with
  lattice simulations
  \cite{Holland:2003jr,Holland:2004sd,Gerhold:2007yb,Gerhold:2010wv,Bulava:2012pb}.}
For larger bare coupling $\lL$, the Higgs mass increases and
approaches a regime of saturation for $\lL\gg 1$. This is reminiscent
to RG-improved perturbation theory, where the bare coupling hits the
Landau pole $\lL\to \infty$ already at a finite cutoff $\Lambda$.

Whereas the Landau pole in perturbation theory in the first place
signals the breakdown of the perturbative expansion, our truncation of
the RG flow does neither rely on perturbative ordering nor require a
weak coupling. Instead, our derivative expansion is organized in terms
of field operators with increasing number of derivatives. In order to
check the convergence of this expansion, we can compare the results
for the Higgs mass to leading order (LO) and next-to-leading order
(NLO) in this expansion. To leading order, we drop the running of the
kinetic terms in \Eqref{eq:trunc} by setting the anomalous dimensions
to zero, $\eta_{\psi,\phi}\to 0$. The resulting Higgs masses are
plotted as dashed lines in Fig.~\ref{fig:lambdamap}. We observe that
the difference to the NLO result (solid lines) is rather small for the
lower Higgs mass bound for $\lL\to 0$; even for the largest accessible
couplings, we observe a maximum deviation of 10\%, confirming that the
derivative expansion constitutes a satisfactory approximation for our
purpose for the whole range from weak to strong coupling.

\begin{figure}
{
\centering\includegraphics[width=8.5cm]{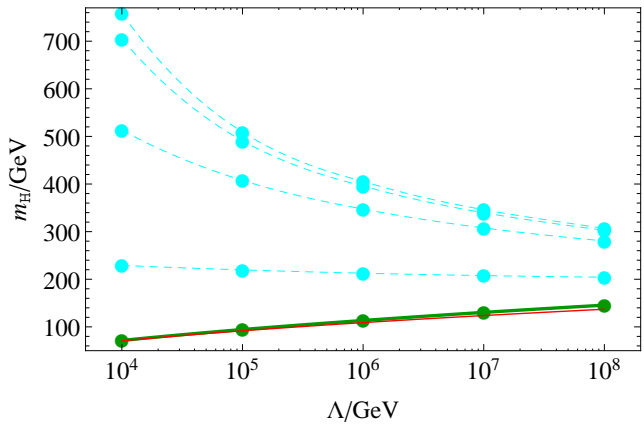}
} 
\caption{Higgs mass bounds versus cutoff $\Lambda$. The thick green/solid
  line denotes the lower bound for the Higgs mass derived within the
  class of bare $\phi^4$ potentials. The thin red/solid line shows the
  lower bound as derived within mean-field approximation. The
  turquois/dashed lines mark upper bounds if the bare scalar coupling
  is allowed to start maximally from $\lL=1,10,50,100$ from bottom to top,
  respectively. An artificial restriction to the perturbative domain
  $\lL\lesssim 1$ underestimates the upper bound by a factor 
  $\gtrsim\mathcal{O}(1)$.  }
\label{fig:boundsphi4}
\end{figure}

Furthermore, we study the convergence of the polynomial expansion of
the scalar potential in Fig.~\ref{fig:lambdamap}. To lowest nontrivial
order $\Np=2$ (green lines with squares), we obtain already a complete
picture of the physics of Higgs mass bounds. For the next order
$\Np=4$ (blue lines with circles), though the upper Higgs mass bound
is already approached for smaller bare couplings $\lL$, the value of
the upper bound changes by at most 5\%. For even higher
orders, the corresponding results lie on top of the $\Np=4$
curves. Within our numerical accuracy we find no significant
difference for $\Np=4,6,8$.

In Fig.~\ref{fig:boundsphi4}, we show the resulting Higgs mass bounds, arising
within the class of $\phi^4$ bare potentials. The thick solid/green line
characterizes the lower bound resulting from the RG flow for a wide range of
cutoffs $\Lambda=10^4 \dots 10^{8}$GeV. Also shown is the lower bound as
derived within the mean-field approximation in the previous section (thin
solid/red line), which
neglects the running of the Yukawa coupling, of the anomalous dimension, and
RG improvement of the scalar potential.  In the full flow, we observe
nontrivial cancelations among these terms, such that the mean-field result
represents a surprisingly good approximation over a wide range of cutoff
scales. The turquois/dashed lines depict upper bounds for the Higgs mass for
bare couplings $\lL=1,10,50,100$, respectively. In particular, we find that if
we limited ourselves to a perturbative domain, choosing $\lL=1$, we would
artificially underestimate the upper bound by a factor
$\gtrsim\mathcal{O}(1)$.

\subsection{Generalized bare potentials}
\begin{figure}
{
\centering\includegraphics[width=8.5cm]{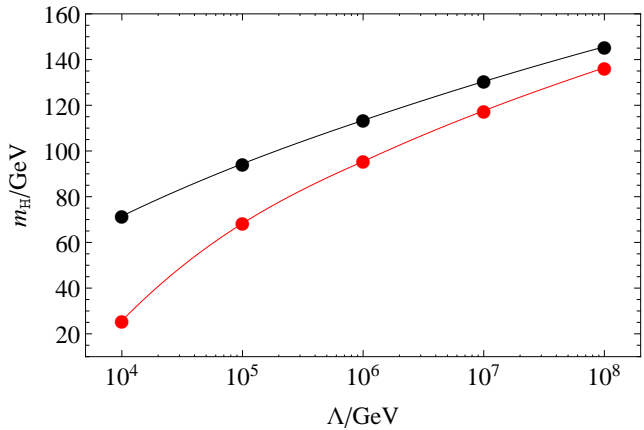}
}
\caption{Higgs mass versus cutoff $\Lambda$. The black
  line denotes the lower bound for the Higgs mass derived within the
  class of bare $\phi^4$ potentials. The red line shows how we can construct 
  Higgs masses below the lower bound by giving up the restriction to quartic 
  bare potentials. The masses are derived for $\lambda_2=-0.1$ and $\lambda_3=3$.}
\label{fig:negphi4}
\end{figure}

Let us now study extensions of the initial bare potential beyond the
$\phi^4$-type.  Motivated by the results of the mean-field
approximation, we concentrate on potentials with a negative
$\lambda_{2,\Lambda}$ where the UV stability is guaranteed by a
positive $\lambda_3\phi^6$. It is possible to construct bare
potentials which give rise to Higgs masses below the lower bound
within the class of $\phi^4$ bare potentials, similar to the
mean-field approach. Fig.~\ref{fig:negphi4} shows the lower bound
within $\phi^4$ theory (black solid line) in comparison to Higgs mass
values for an example flow which starts with
$\lambda_{2,\Lambda}=-0.1$ and $\lambda_{3,\Lambda}=3$ in the UV (red
solid line). This example clearly illustrates that the lower bound
within $\phi^4$-like initial potentials does no longer hold, if higher
dimensional operators are also permitted.

This phenomenon can be understood from the RG flow itself: first we
note that in both cases ($\phi^4$-like as well as the beyond-$\phi^4$
example above) the flow starts in the symmetric regime. In the
beyond-$\phi^4$ example, the quartic coupling $\lambda_2$ runs quickly
to positive values, whereas $\lambda_3$ becomes very small as is
expected in the vicinity of the Gau\ss ian fixed point. As a
consequence, this particular system flows back into the class of
$\phi^4$-type potentials. The decisive difference, however, is that
the scale $k_{\text{GFP}}$ where the system is again near the
Gau\ss{}ian fixed point is now lower than the initial UV scale
$\Lambda$. Loosely speaking, some ``RG time'' is required to run from
the beyond-$\phi^4$ form of the potential back to the $\phi^4$
Gau\ss{}ian type. 

From another viewpoint, the RG flow can map an initial bare action with
$\lambda_2<0$ and $\lambda_3>0$ at an initial UV scale $\Lambda$ to a theory
with $\lambda_2\geq 0$ and $\lambda_3\approx 0$ at a smaller scale
$k_{\text{GFP}}<\Lambda$. Therefore, the red curve (beyond-$\phi^4$) in
Fig.~\ref{fig:negphi4} can also be viewed as a horizontally displaced version
of the black curve ($\phi^4$-like) to effectively larger cutoff values. We
emphasize that the present example has neither been specifically designed or
fine-tuned, nor does it represent an exhaustive study of admissible initial
potentials. A wide range of beyond-$\phi^4$ potentials initiating the flow at
$\Lambda$ leads to Higgs masses below the bound of the $\phi^4$-type
class. Still, the mechanism observed above starting from stable potentials with
$\lambda_2<0$ and globally stabilizing higher-order terms appears rather
generic. We have also checked for more involved initial conditions that
the results for the Higgs masses do not change for higher-order $N_P\geq 4$ 
polynomial expansions of the scalar potential.

In fact, the influence of higher dimensional operators has also been studied
in recent lattice simulations in a chiral Higgs-Yukawa model
\cite{Bulava:2012pb}, by adding a positive $\lambda_3 \phi^6$ term to the bare
potential. No lowering of the Higgs mass bound has been observed in this
study. This is indeed in agreement with our observations, because merely
adding this term has barely any effect on the Higgs mass bound and rather
leads to an increase of the Higgs mass. Our mechanism for lowering the mass
bound works particularly well for initial potentials with $\lambda_2< 0$. In
other words, the $\lambda_2< 0$ deformation requires a comparatively long RG
time to run the potential back to the $\phi^4$ Gau\ss{}ian type. A lattice
study with such (or even more general) bare potentials could hence put our
mechanism to test.

Having put the significance of the lower bound of the Higgs mass derived for
$\phi^4$-type bare potentials into perspective, let us address the issue of
stability: while the standard approach to vacuum stability in the present
simple model based on RG-improved perturbation theory has been questioned by
lattice simulations \cite{Holland:2003jr,Holland:2004sd} and functional RG
methods \cite{Branchina:2005tu} (in turn critically assessed by
\cite{Einhorn:2007rv}), a full stability analysis would require to follow the
RG flow of arbitrary physically admissible initial potentials. In particular,
the RG evolution of potentials with multiple local minima would have to be
dealt with quantitatively. While this is indeed possible with appropriate
numerical solvers
\cite{Gneiting:2005,Hofling:2002hj,Adams:1995cv,Bohr:2000gp,Fischbacher:2012ib},
we here confine ourselves to the validity region of the polynomial expansion
of the effective potential about a local minimum.

Since high-order polynomials typically have multiple local minima, we have to
estimate the radius of convergence of our expansion in field space. A new
local minimum showing up within this convergence region could then be
interpreted as a signature of instability. If such minima only occur outside
the convergence radius, we consider them as an artifact of the polynomial
expansion.

\begin{figure}[t]
{
\centering\includegraphics[width=8.5cm]{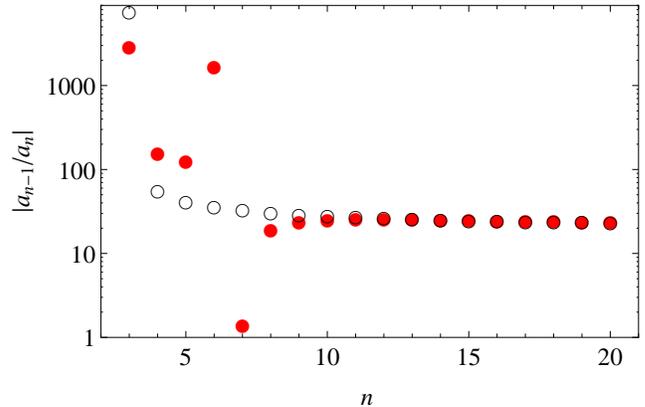}
} 
\caption{Estimate for the radius of convergence in units of $10^3\text{GeV}^2$
  of the polynomial expansion of the effective potentials in terms of the
  absolute values of the ratios of expansion coefficients. The red filled
  circles are derived for a theory which starts at $\Lambda=10^7\,$GeV with
  all couplings set to zero apart from the mass term. The black empty circles
  are for the case $\Lambda=10^7\,$GeV and $\lambda_{2}=1$ and $\lambda_{n}=0$
  ($n\geq 3$).}
\label{fig:ratioofcouplings}
\end{figure}

A rough estimate for the radius of convergence is given by comparing
the quotients of successive couplings $\lambda_n/\lambda_{n+1}$ for
large $n$ in the infrared. In practice we solve the system of coupled
differential equations for $N_P=20$, switching back to dimensionful
quantities at a scale where the flows are frozen out, e.g. $U_k=u_k
k^4=\sum_n a_n({Z_\phi}\rho- v^2/2)^n$ with
$a_n=\frac{\lambda_n}{n!}k^{4-2n}$, and computing the dimensionful
radius of convergence by comparing $\frac{a_n(k)}{a_{n+1}(k)}$ for
$k\rightarrow 0$.  The results expressed in units of $10^3$GeV for
various initial conditions 
%
are plotted in Fig. \ref{fig:ratioofcouplings}.

Our primary observation is that this estimate for the radius of
convergence appears to stabilize at a universal value rather
independent of the chosen initial conditions. The resulting value near
$\simeq 23 000$GeV$^2$ is of the order of the vacuum expectation value
$v^2/2=30258\,$GeV$^2$ for large $n$.  We still observe a slight drift
in our data even at high order, which might be due to the fact that
the inner region of the effective potential owing to its convexity
cannot be resolved within a polynomial expansion as a matter of
principle.  Restricting the field amplitudes to values of the order of
the ratio of the highest couplings in the truncation,
$Z_\phi\rho_{\text{max}}\simeq (v^2/2) + |\frac{a_{N_P-1}}{a_{N_P}}|$,
we find in all studied cases that the effective potential is a convex
monotonically rising function in the outer region ($\phi > v$). No
evidence for an instability within this radius of convergence is
found.

These observations agree with solutions of the RG flow for the full
effective potential beyond the polynomial expansion as worked out in
\cite{Gneiting:2005} using pseudo-spectral methods (Chebyshev
expansion). Both methods lead to equivalent results for both, the
Higgs mass bounds for $\phi^4$-type initial potentials as well as the
absence of any indication for an instability.


\section{Conclusions}    

We have determined Higgs mass bounds in a simple Higgs-Yukawa toy model
sharing some similarities with the standard model Higgs--top-quark sector. Our
study is based on the functional renormalization group which can keep track of
threshold phenomena, has better access to strong coupling regimes and
automatically accounts for ``RG improvement''. 

In agreement with the standard literature, the existence of an upper Higgs
mass bound is a consequence of triviality of the scalar sector. As such, it is
inherently non-universal. In this work, we have also emphasized the
non-universality of the lower Higgs mass bound. In addition to the
regularization scheme dependence which the lower bound shares with the upper
bound in any UV incomplete theory, we have discovered that the lower mass
bound can depend sensitively on the microscopic details of the bare effective
potential for the Higgs field. 

This observation does not contradict Wilsonian renormalizability arguments
stating that IR observables should be independent of the details of the UV
theory. The reason is that a Higgs mass bound given in the form
$m_{\text{H,bound}}=m_{\text{H,bound}}(\Lambda)$ as a function of the UV
cutoff $\Lambda$ does not constitute a pure IR observable. By contrast, it
should be understood as a mapping of initial conditions at the microscopic UV
scale onto the set of possible IR observables. As the UV initial conditions are
typically not accessible by low energy measurements, they are unconstraint. A
statement about Higgs mass bounds therefore often goes along with (typically
only implicit) constraints on the UV initial conditions, i.e., bare actions or
bare potentials. 

In the conventional discussions of Higgs mass bounds, the IR measured
observables are taken from experiment and the RG flow is run to higher
scales. This procedure lacks any control over RG irrelevant operators, as
their influence on the IR observables is exponentially small. Therefore, their
high-energy behavior is simply ignored or implicitly fixed by computational
recipes such as RG-improved perturbation theory. Latest results along this
line of reasoning show that the measured mass of the Higgs boson is close to
the ``vacuum stability'' bound or even in the ``metastable region'' (with the
biggest uncertainty arising from the exact value of the top mass, to be
specified in an appropriate scheme)
\cite{Degrassi:2012ry,Alekhin:2012py,Masina:2012tz,Buttazzo:2013uya}. From
this viewpoint, the fact that the Higgs mass together with the whole standard
model is close to a phase transition is a remarkable result of the LHC,
requiring an explanation of this ``near-criticality'' property
\cite{Buttazzo:2013uya}. Since this running-up of the perturbative RG cannot
access the large field regime, where a new vacuum is expected to occur, a full
resolution of this near-criticality puzzle either requires nonperturbative
complements or even calls for beyond-standard-model explanations. 

Our results offer a different viewpoint: as we have hardly any information
about the bare action at an initial scale $\Lambda$, bounds on particle masses
can only arise from the mapping of {\it all admissible} bare initial conditions
onto the IR observables as is provided by the RG. Of course, the resulting
bounds will depend on the criteria of {\it admissibility} which we may
impose. In this work, we have demonstrated that strict Higgs mass bounds arise
if we restrict the initial conditions to $\phi^4$-type potentials. We
emphasize, however, that this restriction is somewhat arbitrary: it cannot be
justified by Wilsonian renormalizability arguments, as they simply do not
apply to bare actions. Hence, if we lift this artificial restriction, we can
easily discover initial conditions that lead to Higgs masses substantially
smaller than the Higgs mass bound within the $\phi^4$ class. This is already
the case for initial potentials with comparatively small higher-order
operators. Nonperturbatively large deformations of the initial potential are
not required. 

From this viewpoint, the near-criticality property of the standard model
remains nevertheless remarkable, as it may provide for a first handle on the
microscopic action at some high (GUT-like or Planck) scale that has to emerge
from an underlying theory (``a UV completion''). The top-down analog of this
reasoning has been used in a model with asymptotically safe gravity that
predicted the value of the Higgs mass \cite{Shaposhnikov:2009pv} (see also
\cite{Bezrukov:2012sa}), based on the fact that asymptotically safe gravity
interactions are likely to put the Higgs mass onto its ``conventional'' lower
bound. Already earlier, arguments for putting the standard model onto this
conventional lower bound lead to similar predictions \cite{Froggatt:1995rt}.

By contrast, if the Higgs mass turns out to lie below this conventional lower
bound, this may not be a sufficient reason for concern regarding vacuum
stability or metastability. Stability might simply be provided by higher-order
operators in the initial bare action. Rather generically, we find that models
with a negative $\lambda_{2,\Lambda}$ being stabilized by higher-order
operators yield Higgs mass values below the conventional lower bound. Of
course, the presence and magnitude of these higher order operators eventually
has to be explained by a (more) UV complete underlying theory.  In fact,
models with a negative $\lambda_{2,\Lambda}$ have recently been discussed from
a string-theory perspective \cite{Hebecker:2013lha}. A UV complete example for
models with a potentially smaller Higgs mass has recently been given within
pure quantum field theory in the context of an asymptotically safe gauged
Higgs-Yukawa model \cite{Gies:2013pma}.

\section*{Acknowledgments}

We thank Tobias Hellwig, Karl Jansen, Stefan Lippoldt, Axel
Maas, Jan Pawlowski and Luca Zambelli for interesting and enlightening
discussions. HG and RS acknowledge support by the DFG under grants
GRK1523, Gi 328/5-2 (Heisenberg program).

\appendix

\section{Threshold functions}
\label{appA}

In this work, we use the linear regulator which is optimized for the present
truncation \cite{Litim:2001up}. For the bosonic modes, this regulator is given
by
\begin{align*}
R_k(p)= Z_{\phi,k} p^2 \, r(p^2/k^2) = Z_{\phi,k}(k^2-p^2)\theta(k^2-p^2).
\end{align*}
The corresponding chirally symmetric fermionic regulator $R_k(p)=Z_{\psi,k} \fss{p}
r_{\text{F}}(p^2/k^2)$ is chosen such that $p^2(1+r)=p^2(1+r_{\text{F}})^2$. 
For reasons of completeness, we list the threshold functions appearing in the
main text, which can be analytically computed for the linear regulator as a
result of the corresponding momentum integrations:
\begin{align*}
 l_n^d(\omega;\eta_{\phi}) &= \frac{2(\delta_{n,0}+n)}{d}\frac{1-\frac{\eta_{\phi}}{d+2}}{(1+\omega)^{n+1}},
 \\
 l_{0}^{(F)d}(\omega;\eta_{\psi}) &= \frac{2(\delta_{n,0}+n)}{d} \frac{1-\frac{\eta_{\psi}}{d+1}}{(1+\omega)^{n+1}},
 \\
 l_{n_1,n_2}^{(FB)d}(\omega_1,\omega_2;\eta_{\psi},\eta_{\phi}) &= \frac{2}{d}\frac{1}{(1+\omega_1)^{n_1}}\frac{1}{(1+\omega_2)^{n_2}} \\
 &\hspace{-0.45cm}\times \left[ \frac{n_1 \left(1-\frac{\eta_{\psi}}{d+1}\right)}{1+\omega_1} 
         + \frac{n_2 \left(1-\frac{\eta_{\phi}}{d+2}\right)}{1+\omega_2}
       \right], \\
 m_{n_1,n_2}^d(\omega_1,\omega_2;\eta_{\phi}) &= \frac{1}{(1+\omega_1)^{n_1}(1+\omega_2)^{n_2}},\\
 m_2^{(F)d}(\omega;\eta_{\psi}) &= \frac{1}{(1+\omega)^4},
%
 \\
 m_4^{(F)d}(\omega;\eta_{\psi}) &= \frac{1}{(1+\omega)^4} + \frac{1-\eta_{\psi}}{d-2}\frac{1}{(1+\omega)^3} \\ 
 &\quad - \left( \frac{1-\eta_{\psi}}{2d-4} + \frac{1}{4} \right) \frac{1}{(1+\omega)^2},
 \\
 m_{n_1,n_2}^{(FB)d}(\omega_1,\omega_2;\eta_{\psi},\eta_{\phi}) &= \frac{1-\frac{\eta_{\phi}}{d+1}}{(1+\omega_1)^{n_1}(1+\omega_2)^{n_2}}.
\end{align*}
These threshold functions agree with those given in \cite{Hofling:2002hj}.

\end{document}